# Nonlinear anomalous Hall effect for Néel vector detection

Ding-Fu Shao,[1,*,†] Shu-Hui Zhang,[2,*,‡] Gautam Gurung,[1] Wen Yang,[3] and Evgeny Y. Tsymbal[1,§]

[1] *Department of Physics and Astronomy & Nebraska Center for Materials and Nanoscience, University of Nebraska, Lincoln, Nebraska 68588-0299, USA*
[2] *College of Science, Beijing University of Chemical Technology, Beijing 100029, People's Republic of China*
[3] *Beijing Computational Science Research Center, Beijing 100193, People's Republic of China*

Antiferromagnetic (AFM) spintronics exploits the Néel vector as a state variable for novel spintronic devices. Recent studies have shown that the field-like and antidamping spin-orbit torques (SOT) can be used to switch the Néel vector in antiferromagnets with proper symmetries. However, the precise detection of the Néel vector remains a challenging problem. In this letter, we predict that the nonlinear anomalous Hall effect (AHE) can be used to detect the Néel vector in most compensated antiferromagnets supporting the antidamping SOT. We show that the magnetic crystal group symmetry of these antiferromagnets combined with spin-orbit coupling produce a sizable Berry curvature dipole and hence the nonlinear AHE. As a specific example, we consider half-Heusler alloy CuMnSb, which Néel vector can be switched by the antidamping SOT. Based on density functional theory calculations, we show that the nonlinear AHE in CuMnSb results in a measurable Hall voltage under conventional experimental conditions. The strong dependence of the Berry curvature dipole on the Néel vector orientation provides a new detection scheme of the Néel vector based on the nonlinear AHE. Our predictions enrich the material platform for studying non-trivial phenomena associated with the Berry curvature and broaden the range of materials useful for AFM spintronics.

Spintronics is a research field which studies the active control and detection schemes of the spin degrees of freedom in solid-state systems [1]. Over decades, novel phenomena have been discovered in the variety of ferromagnet-based layered structures, forming the core elements for spintronic applications. Recently, the effort in the field has been deployed to enhance the device switching speed and reduce power consumption. In this regard, antiferromagnets are outstanding candidates to replace the widely used ferromagnets in the next generation of spintronic applications, due to their robustness against magnetic perturbations, absence of stray fields, and ultrafast dynamics [2,3,4,5,6].

Recent finding have shown that the field-like spin-orbit torque (SOT) can be used to control the Néel vector in antiferromagnets with $\hat{P}\hat{T}$ symmetry (where $\hat{P}$ is the space inversion symmetry and $\hat{T}$ is the time reversal symmetry) [7,8,9], such as CuMnAs [10], $Mn_2Au$ [11], and $MnPd_2$ [12]. The antidamping SOT can be used for Néel vector switching in globally non-centrosymmetric antiferromagnets such as CuMnSb [8,9,10,13,14]. On the other hand, it is extremely difficult to detect the Néel vector in antiferromagnets using common magnetometers or magnetic resonance techniques due to the absent net magnetization and ultrafast magnetization dynamics [2]. Current experiments exploit the anisotropic magnetoresistance (AMR) effect to detect the Néel vector switching [2,10], where the readout speed is limited by its small magnitude [5]. Accurate detection of the Néel vector orientation is possible using optical methods [15,16,17]. However, the specific requirements of these experimental techniques limit their device application. An efficient electric detection of the Néel vector using a conventional experimental setup would be desirable for a practical AFM spintronic device.

Anomalous Hall effect (AHE) [18,19] is a transport phenomenon driven by the Berry curvature $\bm{\Omega}$, a quantity inherent in the electronic band structure of a material [20,21,22]. Since $\bm{\Omega}$ is odd under time reversal symmetry $\hat{T}$, i.e. $\hat{T}\bm{\Omega}(\bm{k}) = -\bm{\Omega}(-\bm{k})$, where $\bm{k}$ is the wave vector, the integral of $\bm{\Omega}$ over the full Brillouin zone (which determines the AHE) may be non-zero for materials with broken $\hat{T}$ [20]. Since antiferromagnets have broken $\hat{T}$ symmetry, they may support the non-vanishing AHE, which can be used for Néel vector detection. Recently, the AHE has been discovered in non-collinear antiferromagnets (Fig. 1(a)), such as $Mn_3X$ (X = Ga, Ge, Sn, or Ir) [23, 24, 25, 26, 27, 28] and $Mn_3AN$ (A = Ga, Zn, Ag, or Ni) [29,30,31], and collinear antiferromagnets with specific arrangement of non-magnetic atoms in the crystal lattice (Fig. 1(b)), such as $CoNb_3S_6$ [32,33,34]. These compounds hold, however, the net magnetization $\bm{M}$ resulting from the weak canting of the local moments. This is due to no magnetic group symmetry operation $\hat{T}\hat{O}$ ($\hat{O}$ is a crystal space group symmetry operation) enforcing $\bm{M}$ to be zero, as follows from $\hat{T}\hat{O}\bm{M} = -\bm{M}$ [35]. Therefore, these canted antiferromagnets are not truly invisible to the magnetic perturbations. For example, the net magnetization of $Mn_3Ge$ can be rotated by a small magnetic field, leading to reversal of the AHE [26]. To avoid such instabilities, the fully compensated antiferromagnets with $\bm{M} = 0$ are desirable for robust AFM spintronic devices. Zero magnetization requires, however, the $\hat{T}\hat{O}$ symmetry, which prohibits the AHE since $\bm{\Omega}$ is antisymmetric with respect to $\hat{T}\hat{O}$, i.e. $\hat{T}\hat{O}\bm{\Omega}(\bm{k}') = -\bm{\Omega}(\bm{k})$ and $\hat{T}\hat{O}\bm{k}' = \bm{k}$. Thus, it is impossible to have a *linear* AHE in fully compensated antiferromagnets.

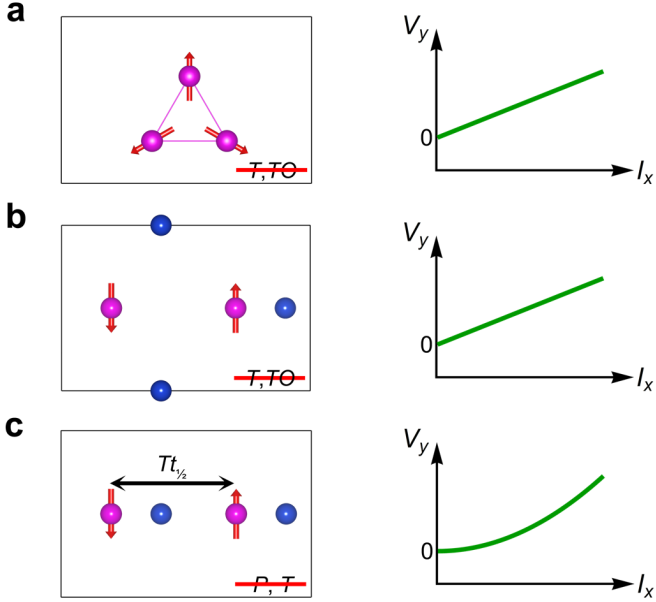

FIG. 1. (a) A non-collinear antiferromagnet with broken $\hat{T}$ and absent $\hat{T}\hat{O}$ symmetry ($\hat{O}$ is a crystal space group symmetry operation), such that $\hat{T}\hat{O}\boldsymbol{\Omega}(\boldsymbol{k}') = -\boldsymbol{\Omega}(\boldsymbol{k})$ (left), resulting in a linear AHE (right). (b) A collinear antiferromagnet with a specific arrangement of non-magnetic atoms in the crystal lattice, in which both $\hat{T}$ and $\hat{T}\hat{O}$ symmetries are absent (left), resulting in a linear AHE (right). (c) A collinear antiferromagnet with broken $\hat{P}$ and $\hat{T}$ symmetries but preserved $\hat{T}\hat{t}_{1/2}$ symmetry (left), resulting in a non-linear AHE (right).

This problem can be solved with a recently discovered *nonlinear* AHE [36,37,38,39,40,41,42]. In contrast to the linear AHE, where the Hall voltage is linear to an applied electric field as found in numerous magnetic (i.e. $\hat{T}$ broken) systems, the nonlinear AHE occurs in second-order response to an electric field as demonstrated for a certain class of non-magnetic (i.e. $\hat{T}$ invariant) materials. The nonlinear AHE requires broken $\hat{P}$ symmetry and arises from the Berry curvature dipole $\boldsymbol{D}$, which generates a net anomalous velocity when the system is in a current-carrying state [36]. So far, however, the nonlinear AHE has been considered only for non-magnetic materials where $\hat{T}$ symmetry is preserved. Extending this concept to AFM materials where $\hat{T}$ symmetry is broken is interesting and desirable, as it would broaden a range of measurable properties useful for AFM spintronics.

In this letter, we predict that the nonlinear AHE does exist in most compensated antiferromagnets supporting the electric control of the Néel vector by antidamping SOT. As a specific example, we consider half-Heusler alloy CuMnSb and demonstrate, based on first-principles density functional theory (DFT) calculations [43], that a polar axis and a combined $\hat{T}\hat{t}_{1/2}$ symmetry (where $\hat{t}_{1/2}$ is translation by half a unit cell) support a sizable nonlinear AHE due to a finite Berry curvature dipole. Moreover, we predict a strong dependence of the Berry curvature dipole and hence the nonlinear AHE on the AFM Néel vector orientation, which can be used for the Néel vector detection in similar compounds.

The Berry curvature dipole density tensor $\boldsymbol{d}$ is defined as $d_{bd} = -\frac{\partial f_0}{\partial k_b}\Omega^d$, where $k_b$ and $\Omega^d$ are Cartesian components of the wave vector and the Berry curvature, respectively, and $f_0$ is the equilibrium Fermi distribution function. It is odd under $\hat{P}$, and therefore the nonlinear AHE in antiferromagnets requires a non-centrosymmetric structure. In fact, the condition is even more stringent and, similar to non-magnetic systems, necessitates the presence of gyrotropic symmetry constraints [36,44,45]. We find that most compensated antiferromagnets supporting the antidamping SOT fulfill the symmetry requirements for the nonlinear AHE. Among the 123 non-centrosymmetric compensated antiferromagnets reported in the Bilbao MagnData database, there are 118 compounds with the magnetic space groups supporting the finite Berry curvature dipole [46,47].

For example, a polar axis and $\hat{T}\hat{t}_{1/2}$ symmetry (Fig. 1(c)) are sufficient for the nonlinear AHE to emerge in an antiferromagnet, due to $\boldsymbol{d}$ being even under the $\hat{T}\hat{t}_{1/2}$ transformation, i.e. $\hat{T}\hat{t}_{1/2}\boldsymbol{d}(\boldsymbol{k}) = \boldsymbol{d}(-\boldsymbol{k})$. This is the case for half-Heusler alloy CuMnSb, an AFM metal supporting the antidamping SOT [8,9,10]. Figs. 2(a,b) show the CuMnSb structure that belongs to the crystal space group $F\bar{4}3m$ [13,14,48]. Below the Néel temperature $T_N = 55$ K, a type-II collinear AFM order emerges, where the magnetic moments of the Mn atoms are parallel within the (111) plane but antiparallel between the successive (111) planes (Fig. 2(a)). The Néel vector is pointed along the [111] direction. This AFM order lowers the symmetry, resulting in the magnetic space group $R_I3c$ [14]. The rhombohedral primitive cell of CuMnSb (Fig. 2(b)) contains three-fold rotation $\hat{C}_3$ around the [111] direction, and three glide mirror reflections $\hat{g}_{\bar{1}10}$, $\hat{g}_{\bar{1}01}$, and $\hat{g}_{0\bar{1}1}$, where $\hat{g}_l = \{\widehat{M}_l | \hat{t}_{1/2}\}$ is mirror symmetry $\widehat{M}_l$ normal to vector $\boldsymbol{l}$ combined with half a unit cell translation $\hat{t}_{1/2}$. In addition, below 34 K the Néel vector is canted towards the [110] direction by $\delta = 11°$ [14,49]. This canting breaks the $\hat{C}_3$, $\hat{g}_{\bar{1}01}$, and $\hat{g}_{0\bar{1}1}$ symmetries, leading to the magnetic space group $C_cc$ in which only $\hat{g}_{\bar{1}10}$ is preserved [14]. Both AFM phases of CuMnSb are polar and contain the $\hat{T}\hat{t}_{1/2}$ symmetry.

First, we investigate the AFM phase of CuMnSb without canting. Assuming the experimental lattice constant $a = 6.075$ Å, we find the calculated magnetic moment of 3.93 μ$_B$/Mn consistent with the experimental value of 3.9 (±0.1) μ$_B$/Mn [14]. Fig. 2(c) shows the calculated band structure. There are six bands crossing the Fermi energy $E_F$, mostly contributed by 3$d$ electrons of Mn [50] (Fig. S1). The four valence bands are very dispersive with the maximum around the Γ point, forming four hole pockets along the Γ-Z direction. The bands at the conduction band minimum are less dispersive, forming electron

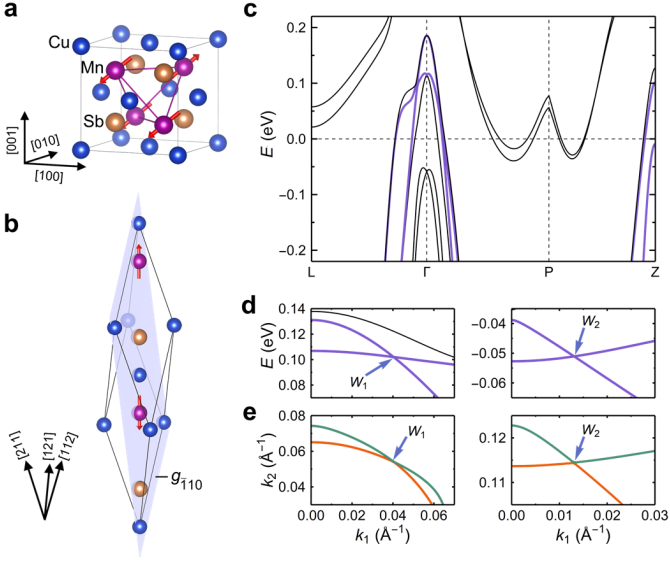

$$\chi^{(0)}_{abc} = \chi^{(2)}_{abc} = -\epsilon_{adc}\frac{e^3\tau D_{bd}}{2\hbar^2(1+i\omega\tau)}, \quad (1)$$

where $\tau$ is the relaxation time, and $D_{bd}$ is the Berry curvature dipole defined as

$$D_{bd} = \int \frac{d^3k}{(2\pi)^3}d_{bd} = -\int \frac{d^3k}{(2\pi)^3}\sum_n \frac{\partial E_{nk}}{\partial k_b}\Omega^d_{nk}\frac{\partial f_0}{\partial E_{nk}}. \quad (2)$$

Here $E_{nk}$ is the energy of the $n$-th band at the $k$ point. The Berry curvature of the $n$-th band is given by [20,22]:

$$\Omega^d_{nk} = i\epsilon_{abd}\sum_{m\neq n}\frac{\langle n|\frac{\partial H}{\partial k_a}|m\rangle\langle m|\frac{\partial H}{\partial k_b}|n\rangle}{(E_{nk}-E_{mk})^2}. \quad (3)$$

The factor $\frac{\partial E_{nk}}{\partial k_b}$ is odd under both $\hat{P}$ and $\hat{T}$ symmetries and $\Omega^d_{nk}$ is odd under $\hat{T}$ and even under $\hat{P}$ [22]. As a result, $d_{bd}$ in Eq. (2) is even with respect to $\hat{T}$, leading to non-zero $D_{bd}$ and a finite nonlinear AHE in a non-centrosymmetric system without magnetism. In an antiferromagnet like CuMnSb, the preserved $\hat{T}\hat{t}_{1/2}$ symmetry plays the same role as $\hat{T}$ on $\frac{\partial E_{nk}}{\partial k_b}$ and $\Omega^d_{nk}$, and the polar axis ensures a nonzero $\boldsymbol{D}$. Therefore, a finite $D_{bd}$ can be also expected in such an antiferromagnet.

There is only one independent element of the $\boldsymbol{D}$ tensor in the AFM phase without canting (Table SII) [43]. For definiteness, we consider $D_{xz}$ directly related to $J_y = \chi_{yxx}\mathcal{E}_x^2$, a transverse nonlinear Hall current along the $y$ ([010]) direction produced by a longitudinal electric field along the $x$ ([100]) direction. Figs. 3(a,b) show the projection of $D_{xz}$ on the $k_1$-$k_2$ plane, at $E = 0.102$ eV (Fig. 3(a)) and $E = -0.051$ eV (Fig. 3(b)), which is obtained by integration $\int \frac{dk_3}{2\pi}d_{xz}$. According to Eq. (2), $D_{xz}$ is a Fermi-surface property, and thus only the Fermi pockets contribute to $D_{xz}$. We find that the contribution by the Fermi pockets from the conduction bands is negligible compared to those from the valence bands. This is due to the valence bands near their maximum having stronger dispersion, which leads to larger velocity $v_x = \frac{\partial E_{nk}}{\partial k_x}$ (Fig. 2(c)). On the other hand, the gaps between the valence bands are very small near $E_F$, which leads to larger $\Omega^z_{nk}$ according to Eq. (3). Therefore, as seen from the energy dependence of $D_{xz}$ in Fig. 3(c), the magnitude of $D_{xz}$ increases with energy decreasing, due to the increase of the volume of the central Fermi surfaces from the valence bands. The calculated $D_{xz}$ is about $-0.048$ at $E_F$, and can be enhanced to $-0.080$ by proper doping (Fig. 3(c)). These values are comparable to those obtained for non-magnetic metals [38,39].

We note that the contributions of the Weyl fermions $W_1$ and $W_2$ to $D_{xz}$ are different. The states near $W_1$ dominate $D_{xz}$ around $E = 0.102$ eV, as shown in Figs. 3(a) and S4, leading to the anomaly in the $D_{xz}$ curve at this energy (Fig. 3(c)). On the other hand, there is no pronounced contributions from $W_2$ (Fig.

FIG. 2. (a,b) The collinear AFM magnetic structure of CuMnSb shown in a conventional cubic unit cell (a) and a rhombohedral primitive cell (b). Red arrows denote the magnetic moments of Mn. The light blue plane denotes the glide plane $\hat{g}_{\bar{1}10}$. (c) The band structure of CuMnSb near $E_F$. (d) The band structures close to Weyl points $W_1$ with $k_2 = 0.054$ Å$^{-1}$ and $k_3 = 0.004$ Å$^{-1}$ (left) and $W_2$ with $k_2 = 0.115$ Å$^{-1}$ and $k_3 = 0.081$ Å$^{-1}$ (right). Here $k_1$, $k_2$, and $k_3$ are along the $[\bar{1}10]$, $[\bar{1}\bar{1}2]$, and $[111]$ directions in the cubic lattice. The purple lines in (c,d) denote the two bands forming Weyl points. The horizontal dashed line indicates the Fermi energy. (e) The Fermi surfaces at $k_3 = 0.004$ Å$^{-1}$ and $E = 0.102$ eV (left) and at $k_3 = 0.081$ Å$^{-1}$ and $E = -0.051$ eV (right). The Weyl points locate at the intersection points of the Fermi surfaces.

pockets near the edges of the top and bottom surfaces of the Brillouin zone (Fig. S1). There are multiple crossings and anticrossings in the band structure near $E_F$. For example, in Figs. 2(c,d) we show the band crossings by the second and third valence bands (purple lines). As seen from Fig. 2(d), there are two Weyl points at $E = 0.102$ eV ($W_1$) and $E = -0.051$ eV ($W_2$). The band crossings close to the Weyl points are strongly tilted. As is evident from in Fig. 2(e), $W_1$ and $W_2$ are located at the touching points of the two Fermi pockets. This indicates that $W_1$ and $W_2$ are type-II Weyl fermions [51,52]. By application of $\hat{T}\hat{t}_{1/2}$, $\hat{C}_3$, and glide symmetry transformations to $W_1$ and $W_2$, we obtain six pairs of $W_1$ and six pairs of $W_2$ Weyl fermions located in the central part of the Brillouin zone (Fig. S1).

Next, we discuss the nonlinear Hall response. Electric field $\boldsymbol{E}_c = \mathrm{Re}\{\mathcal{E}e^{i\omega t}\}$ of amplitude $\mathcal{E}$ and frequency $\omega$ produces nonlinear current $J_a = \mathrm{Re}\{J_a^{(0)} + J_a^{(2)}e^{i2\omega t}\}$, where $J_a^{(0)} = \chi^{(0)}_{abc}\mathcal{E}_b\mathcal{E}_c^*$ describes the rectified current and $J_a^{(2)} = \chi^{(2)}_{abc}\mathcal{E}_b\mathcal{E}_c^*$ describes the second harmonic current. For a system with time reversal symmetry, the response coefficients are [36]:

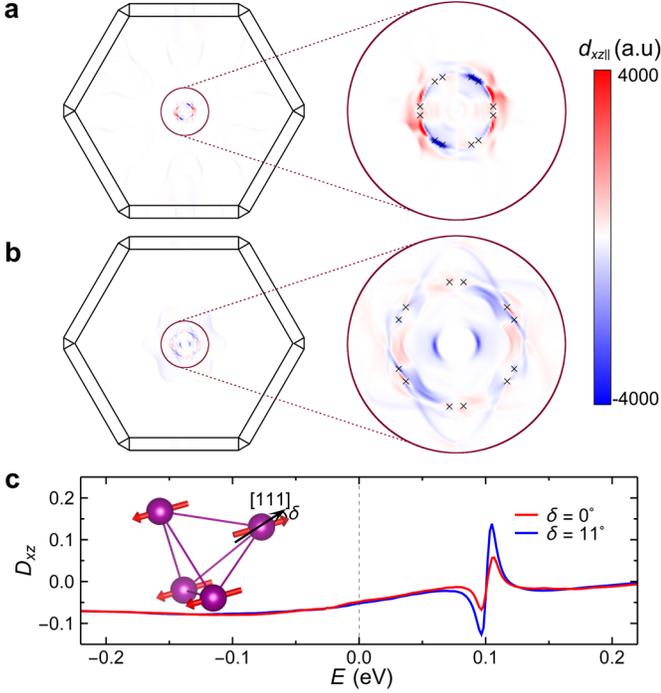

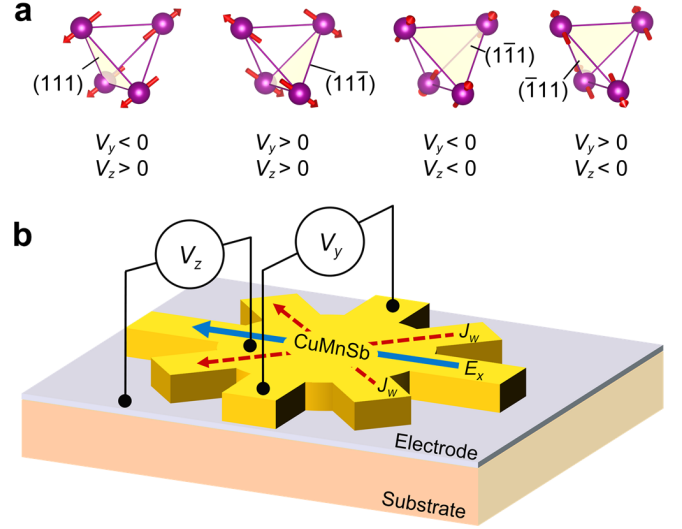

FIG. 3. (a,b) The projections of the Berry curvature dipole on $k_1$-$k_2$ plane at $E = 0.102$ eV (a) and $E = -0.051$ eV (b). Black cross symbols denote position of the Weyl fermions. (c) The $D_{xz}$ as a function of energy for the AFM phase without (red line) and with (blue line) canting of the Néel vector. The inset schematically shows the canting.

FIG. 4. (a) The signs of the Hall voltages $V_y$ and $V_z$ for different AFM orders of CuMnSb. (b) A spin-orbit torque device where the AFM orders of CuMnSb are switched using the antidamping-like SOT generated by the writing current $J_w$.

3(b)) due to the weak tilting of the Weyl cones [43], which suppresses the anomaly in the $D_{xz}$ curve at $E = -0.051$ eV (Fig. 3(c)). We note that the calculated $D_{xz}$ at $E_F$ is independent of $W_1$ and $W_2$, since $D$ is the Fermi surface property, and $W_1$ and $W_2$ lie away from $E_F$. The suitable symmetry, dispersive band structure, and strong spin-orbit coupling are sufficient to produce a sizable $D$ at $E_F$.

At low temperature, the Néel vector is tilted towards the [110] direction by $\delta = 11°$ [14]. This weak canting leads to tiny changes of the band structure (Fig. S2). Nevertheless, breaking the $\hat{C}_3$, $\hat{g}_{\bar{1}01}$, and $\hat{g}_{0\bar{1}1}$ symmetries alters the number and the positions of the Weyl fermions (Fig. S2). The energy dependence of $D_{xz}$ in this AFM phase is generally similar to that in the AFM phase without canting, except sharper peaks above $E_F$ due to the different distribution of the Weyl points (Figs. 3(c) and S2). Moreover, the change of symmetry influences the $D$ tensor [43], resulting in different magnitudes of $D_{xy}$ and $D_{xz}$ (Table SII). This difference can be further enhanced by stronger canting. For example, for the canting angle of $\delta = 90°$ (similar to that in a half Heusler alloy GdMnBi [53]), we find $D_{xy} = 0.072$ being much larger than $D_{xz}$ (Table SII). Since $D_{xy}$ is related to the Hall voltage $V_z$, the large difference between $D_{xy}$ and $D_{xz}$ leads to different nonlinear Hall responses in the y and z directions, which can be used to detect Néel vector canting.

The sensitivity of the Berry curvature dipole to the Néel vector orientation implies that the nonlinear AHE can be used to detect it. The type-II AFM order with the Néel vector normal to the (111) planes in CuMnSb is energetically identical to that normal to the $(\bar{1}11)$, $(1\bar{1}1)$, or $(11\bar{1})$ planes. These orders are expected to have different signs of the transverse and vertical Hall voltage (Fig. 4(a)). For example, switching the Néel vector from (111) to $(11\bar{1})$ is equivalent to the $C_2$ rotation around the [001] direction. This operation changes sign of $D_{xz}$ but does not affect $D_{xy}$. Therefore, $V_y$ is reversed and $V_z$ does not change by this switching. Fig. 4(b) schematically shows a SOT device for such switching. The Néel vector is switched by the antidamping SOT using current $J_w$ (usually $\sim 10^7$-$10^8$ A/cm$^2$). The switched Néel vector can be detected by measuring the Hall voltage $V_y$ and $V_z$, as shown in Fig. 4(a). Considering a sample of 100 μm in length and typical reading current $J_x = 5 \times 10^6$ A/cm$^2$, the Hall voltage of $V_y \sim 14 - 90$ μV in the DC limit is estimated [43], which is well within the capacity of experiments.

Importantly, a high frequency electric field can generate a steady rectified component and the second harmonic component in the Hall current. This allows distinguishing the non-linear AHE signal from the noise induced by the input first harmonic electric field. This feature implies the possible measurement of nonlinear AHE generated by high frequency fields such as picosecond pulses using the noncontact technology, which is very promising for efficient ultrafast detection [40, 54, 55].

The non-linear AHE is expected to exist in a broad range of AFM materials, which Néel vector can be controlled by the antidamping SOT. In addition to CuMnSb, AFM metals PdMnTe [56] and $Ca_3Ru_2O_7$ [57, 58, 59] are promising candidates to host a large nonlinear AHE due to the strong spin-orbit coupling induced by the heavy metal elements.

In conclusion, we have predicted that the nonlinear AHE exists in most compensated antiferromagnets supporting the electric control of the Néel vector by antidamping SOT. As an example, we have considered half-Heusler alloy CuMnSb and showed that this antiferromagnet has a large Berry curvature dipole resulting in a sizable nonlinear AHE. The strong dependence of the Berry curvature dipole on the Néel vector orientation provides a new detection scheme of the AFM order, which is useful for AFM spintronics. We hope therefore that our theoretical predictions will motivate experimentalists to explore the nonlinear AHE in antiferromagnets.

**Acknowledgments.** The authors thank Ruichun Xiao and Bo Li for helpful discussions. This work was supported by the National Science Foundation (NSF) through the Nebraska MRSEC program (grant DMR-1420645) and the DMREF program (grant DMR-1629270). S.-H.Z thanks the support of National Science Foundation of China (NSFC Grants No. 11504018 and No. 11774021). Computations were performed at the University of Nebraska Holland Computing Center.

[*] These authors contributed equally to this work.

[†] dfshao@unl.edu

[‡] shuhuizhang@mail.buct.edu.cn

[§] tsymbal@unl.edu

[1] E. Y. Tsymbal and I. Žutić Eds., *Spintronics Handbook: Spin Transport and Magnetism*, 2nd edition (CRC press, 2019).

[2] V. Baltz, A. Manchon, M. Tsoi, T. Moriyama, T. Ono, and Y. Tserkovnyak, Antiferromagnetic spintronics, *Rev. Mod. Phys.* **90**, 015005 (2018).

[3] T. Jungwirth, X. Marti, P. Wadley, and J. Wunderlich, Antiferromagnetic spintronics, *Nat. Nanotech.* **11**, 231 (2016).

[4] T. Jungwirth, J. Sinova, A. Manchon, X. Marti, J. Wunderlich, and C. Felser, The multiple directions of antiferromagnetic spintronics, *Nat. Phys.* **14**, 200 (2018).

[5] J. Železný, P. Wadley, K. Olejník, A. Hoffmann, and H. Ohno, Spin transport and spin torque in antiferromagnetic devices, *Nat. Phys.* **14**, 220 (2018)

[6] L. Šmejkal, Y. Mokrousov, B. Yan, and A. H. MacDonald, Topological antiferromagnetic spintronics, *Nat. Phys.* **14**, 242 (2018).

[7] J. Železný, H. Gao, K. Výborný, J. Zemen, J. Mašek, A. Manchon, J. Wunderlich, J. Sinova, and T. Jungwirth, Relativistic Néel-Order Fields Induced by Electrical Current in Antiferromagnets, *Phys. Rev. Lett.* **113**, 157201 (2014).

[8] J. Železný, H. Gao, A. Manchon, F. Freimuth, Y. Mokrousov, J. Zemen, J. Mašek, J. Sinova, and T. Jungwirth, Spin-orbit torques in locally and globally noncentrosymmetric crystals: Antiferromagnets and ferromagnets, *Phys. Rev. B* **95**, 014403 (2017).

[9] A. Manchon, J. Zelezný, I. M. Miron, T. Jungwirth, J. Sinova, A. Thiaville, K. Garello, and P. Gambardella, Current-induced spin-orbit torques in ferromagnetic and antiferromagnetic systems, *Rev. Mod. Phys.* **91**, 035004 (2019).

[10] P. Wadley, B. Howells, J. Železný, C. Andrews, V. Hills, R. P. Campion, V. Novák, K. Olejník, F. Maccherozzi, S. S. Dhesi, S. Y. Martin, T. Wagner, J. Wunderlich, F. Freimuth, Y. Mokrousov, J. Kuneš, J. S. Chauhan, M. J. Grzybowski, A. W. Rushforth, K. W. Edmonds, B. L. Gallagher, and T. Jungwirth, Electrical switching of an antiferromagnet, *Science* **351**, 587 (2016).

[11] S. Y. Bodnar, L. Šmejkal, I. Turek, T. Jungwirth, O. Gomonay, J. Sinova, A. A. Sapozhnik, H.-J. Elmers, M. Kläui, and M. Jourdan, Writing and reading antiferromagnetic $Mn_2Au$ by Néel spin-orbit torques and large anisotropic magnetoresistance, *Nat. Commun.* **9**, 348 (2018).

[12] D. -F. Shao, G. Gurung, S. -H. Zhang, and E. Y. Tsymbal, Dirac nodal line metal for topological antiferromagnetic spintronics, *Phys. Rev. Lett.* **122**, 077203 (2019).

[13] K. Endo, T. Ohoyama, and R. Kimura, Antiferromagnetism of CuMnSb, *J. Phys. Soc. Jpn.* **25**, 907 (1968).

[14] A. Regnat, A. Bauer, A. Senyshyn, M. Meven, K. Hradil, P. Jorba, K. Nemkovski, B. Pedersen, R. Georgii, S. Gottlieb-Schönmeyer, and C. Pfleiderer, Canted antiferromagnetism in phase-pure CuMnSb, *Phys. Rev. Mater.* **2**, 054413 (2018).

[15] M. J. Grzybowski, P. Wadley, K. W. Edmonds, R. Beardsley, V. Hills, R. P. Campion, B. L. Gallagher, J. S. Chauhan, V. Novak, T. Jungwirth, F. Maccherozzi, and S. S. Dhesi, Imaging current-induced switching of antiferromagnetic domains in CuMnAs, *Phys. Rev. Lett.* **118**, 057701 (2017).

[16] V. Saidl, P. Nemec, P. Wadley, V. Hills, R. P. Campion, V. Novak, K.W. Edmonds, F. Maccherozzi, S. S. Dhesi, B. L. Gallagher, F. Trojanek, J. Kunes, J. Zelezny, P. Maly, and T. Jungwirth, Optical determination of the Néel vector in a CuMnAs thin-film antiferromagnet, *Nat. Photonics* **11**, 91-97 (2017).

[17] Z. Sun, Y. Yi, T. Song, G. Clark, B. Huang, Y. Shan, S. Wu, D. Huang, C. Gao, Z.Chen,M.McGuire,T.Cao,D. Xiao, W.-T. Liu, W. Yao, X. Xu, and S. Wu, Giant nonreciprocal second harmonic generation from layered antiferromagnetc bilayer $CrI_3$, *Nature* **572**, 497 (2019).

[18] R. Karplus and J. M. Luttinger, Hall effect in ferromagnetics, *Phys. Rev.* **95**, 1154 (1954).

[19] N. Nagaosa, J. Sinova, S. Onoda, A. H. MacDonald, and N. P. Ong, Anomalous Hall effect, *Rev. Mod. Phys.* **82**, 1539 (2010).

[20] M. V. Berry, Quantal phase factors accompanying adiabatic changes, *Proc. R. Soc. A* **392**, 45 (1984).

[21] M. -C. Chang and Q. Niu, Berry phase, hyperorbits, and the Hofstadter spectrum: Semiclassical dynamics in magnetic Bloch bands, *Phys. Rev. B* **53**, 7010 (1996).

[22] D. Xiao, M.-C. Chang, and Q. Niu, Berry phase effects on electronic properties, *Rev. Mod. Phys.* **82**, 1959 (2010).

[23] H. Chen, Q. Niu, and A. H. MacDonald, Anomalous Hall effect arising from noncollinear antiferromagnetism, *Phys. Rev. Lett.* **112**, 017205 (2014).

[24] J. Kübler and C. Felser, Non-collinear antiferromagnets and the anomalous Hall effect, *EPL* **108**, 67001 (2014).

[25] S. Nakatsuji, N. Kiyohara, and T. Higo, Large anomalous Hall effect in a non-collinear antiferromagnet at room temperature, *Nature* **527**, 212 (2015).

[26] A. K. Nayak, J. E. Fischer, Y. Sun, B. Yan, J. Karel, A. C. Komarek, C. Shekhar, N. Kumar, W. Schnelle, J. Kübler, C. Felser, and S. P. P. Parkin, Large anomalous Hall effect driven by a nonvanishing Berry curvature in the noncollinear antiferromagnet $Mn_3Ge$, *Sci. Adv.* **2**, e1501870 (2016).

[27] Y. Zhang, Y. Sun, H. Yang, J. Železný, S. P. P. Parkin, C. Felser, and B. Yan, Strong anisotropic anomalous Hall effect and spin Hall effect in the chiral antiferromagnetic compounds $Mn_3X$ (X=Ge, Sn, Ga, Ir, Rh, and Pt), *Phys. Rev. B* **95**, 075128 (2017).

[28] N. Kiyohara, T. Tomita, and S. Nakatsuji, Giant Anomalous Hall Effect in the Chiral Antiferromagnet $Mn_3Ge$, *Phys. Rev. Applied* **5**, 064009 (2016).

[29] G. Gurung, D.-F. Shao, T. R. Paudel, and E. Y. Tsymbal, Anomalous Hall conductivity of non-collinear magnetic antiperovskites, *Phys. Rev. Mater.* **3**, 044409 (2019).

[30] D. Boldrin, I. Samathrakis, J. Zemen, A. Mihai, B. Zou, B. Esser, D. McComb, P. Petrov, H. Zhang, and L. F. Cohen, The anomalous Hall effect in non-collinear antiferromagnetic $Mn_3NiN$ thin films, *Phys. Rev. Materials* **3**, 094409 (2019).

[31] X. Zhou, J. -P. Hanke, W. Feng, F. Li, G.-Y. Guo, Y. Yao, S. Blügel, and Y. Mokrousov, Spin-order dependent anomalous Hall effect and magneto-optical effect in the noncollinear antiferromagnets $Mn_3XN$ with X = Ga, Zn, Ag, or Ni, *Phys. Rev. B* **99**, 104428 (2019).

[32] N. J. Ghimire, A. S. Botana, J. S. Jiang, J. Zhang, Y.-S. Chen, and J. F. Mitchell, Large anomalous Hall effect in the chiral-lattice antiferromagnet $CoNb_3S_6$, *Nat. Commun.* **9**, 3280 (2018).

[33] L. Šmejkal, R. González-Hernández, T. Jungwirth, and J. Sinova, Crystal Hall effect in collinear antiferromagnets, *arXiv*:1901.00445 (2019).

[34] X. Li, A. H. MacDonald, and H. Chen, Quantum anomalous Hall effect through canted antiferromagnetism, *arXiv*:1902.10650 (2019).

[35] H. Chen, T. -C. Wang, D. Xiao, G. -Y. Guo, Q. Niu, and A. H. MacDonald, Manipulating anomalous Hall antiferromagnets with magnetic fields, *arXiv*:1802.03044 (2018).

[36] I. Sodemann and L. Fu, Quantum nonlinear Hall effect induced by Berry curvature dipole in time-reversal invariant materials, *Phys. Rev. Lett.* **115**, 216806 (2015).

[37] Z. Z. Du, C. M. Wang, H. -Z. Lu, and X. C. Xie, Band signatures for strong nonlinear Hall effect in bilayer $WTe_2$, *Phys. Rev. Lett.* **121**, 266601 (2018).

[38] Y. Zhang, Y. Sun, and B. Yan, Berry curvature dipole in Weyl semimetal materials: An *ab initio* study, *Phys. Rev. B* **97**, 041101(R) (2018).

[39] J. I. Facio, D. Efremov, K. Koepernik, J. -S. You, I. Sodemann, and J. van den Brink, Strongly enhanced Berry dipole at topological phase transitions in BiTeI, *Phys. Rev. Lett.* **121**, 246403 (2018).

[40] Q. Ma, S. -Y. Xu, H. Shen, D. MacNeill, V. Fatemi, T. -R. Chang, A. M. Mier Valdivia, S. Wu, Z. Du, C. -H. Hsu, S. Fang, Q. D. Gibson, K. Watanabe, T. Taniguchi, R. J. Cava, E. Kaxiras, H. -Z. Lu, H. Lin, L. Fu, N. Gedik, and Pa. Jarillo-Herrero, Observation of the nonlinear Hall effect under time-reversal-symmetric conditions, *Nature* **565**, 337-342 (2019).

[41] K. Kang, T. Li, E. Sohn, J. Shan, and K. F. Mak, Nonlinear anomalous Hall effect in few-layer $WTe_2$, *Nat. Mater.* **18**, 324 (2019).

[42] J. E. Moore and J. Orenstein, Confinement-Induced Berry Phase and Helicity-Dependent Photocurrents, *Phys. Rev. Lett.* **105**, 026805 (2010).

[43] See Supplemental Material [url] for the calculation methods, electronic structures of CuMnSb, symmetry analysis of the Berry curvature dipole, analysis of the relation between the Weyl points and the Berry curvature dipole, and the estimation of the nonlinear Hall voltage under conventional experimental conditions, which includes Refs. [60-69].

[44] S. S. Tsirkin, P. A. Puente, and I. Souza, Gyrotropic effects in trigonal tellurium studied from first principles, *Phys. Rev. B* **97**, 035158 (2018).

[45] E. J. König, M. Dzero, A. Levchenko, and D. A. Pesin, Gyrotropic Hall effect in Berry-curved materials, *Phys. Rev. B* **99**, 155404 (2019).

[46] S. V. Gallego, J. M. Perez-Mato, L. Elcoro, E. S. Tasci, R. M. Hanson, K. Momma, M. I. Aroyo and G. Madariaga, MAGNDATA: towards a database of magnetic structures. I. The commensurate case, *J. Appl. Cryst.* 49, 1750 (2016).

[47] S. V. Gallego, J. M. Perez-Mato, L. Elcoro, E. S. Tasci, R. M. Hanson, M. I. Aroyo and G. Madariaga, MAGNDATA: towards a database of magnetic structures. II. The incommensurate case, *J. Appl. Cryst.* 49, 1941 (2016).

[48] K. Manna, Y. Sun, L. Müchler, J. Kübler, and C. Felser, Heusler, Weyl, and Berry, *Nat. Rev. Mater.* **3**, 244 (2018).

[49] F. Máca, J. Kudrnovský, V. Drchal, I. Turek, O. Stelmakhovych, P. Beran, A. Llobet, and X. Marti, Defect-induced magnetic structure of CuMnSb, *Phys. Rev. B* **94**, 094407 (2016).

[50] T. Jeong, Ruben Weht, and W. E. Pickett, Semimetallic antiferromagnetism in the half-Heusler compound CuMnSb, *Phys. Rev. B* **71**, 184103 (2005).

[51] N. P. Armitage, E. J. Mele, and A. Vishwanath, Weyl and Dirac semimetals in three-dimensional solids, *Rev. Mod. Phys.* **90**, 015001 (2018).

[52] A. A. Soluyanov, D. Gresch, Z. Wang, Q. S. Wu, M. Troyer, X. Dai, and B. A. Bernevig, Type-II Weyl semimetals, *Nature* **527**, 495-498 (2016).

[53] R. A. Müller, N. R. Lee-Hone, L. Lapointe, D. H. Ryan, T. Pereg-Barnea, A. D. Bianchi, Y. Mozharivskyj, and R. Flacau, Magnetic structure of GdBiPt: A candidate antiferromagnetic topological insulator, *Phys. Rev. B* **90**, 041109(R) (2014).

[54] Hiroki Isobe, Su-Yang Xu, and Liang Fu, High-frequency rectification via chiral Bloch electrons, *arXiv*:1812.08162 (2018).

[55] O. Matsyshyn, I. Sodemann, Nonlinear Hall acceleration and the quantum rectification sum rule, *Phys. Rev. Lett.* **123**, 246602 (2019).

[56] R. B. Helmholdt and K. H. J. Buschow, A neutron diffraction and magnetization study of PdMnTe, *J. LessCommon Met.* **123**, 169 (1986).

[57] Y. Yoshida, S.-I. Ikeda, H. Matsuhata, N. Shirakawa, C. H. Lee, and S. Katano, Crystal and magnetic structure of $Ca_3Ru_2O_7$, *Phys. Rev. B* **72**, 054412 (2005).

[58] Wei Bao, Z. Q. Mao, Z. Qu, and J. W. Lynn, Spin Valve Effect and Magnetoresistivity in Single Crystalline $Ca_3Ru_2O_7$, *Phys. Rev. Lett.* **100**, 247203 (2008).

[59] B. Bohnenbuck, I. Zegkinoglou, J. Strempfer, C. Schüßler-Langeheine, C. S. Nelson, Ph. Leininger, H.-H. Wu, E. Schierle, J. C. Lang, G. Srajer, S. I. Ikeda, Y. Yoshida, K. Iwata, S. Katano, N. Kikugawa, and B. Keimer, Magnetic structure and orbital state of $Ca_3Ru_2O_7$ investigated by resonant x-ray diffraction, *Phys. Rev. B* **77**, 224412 (2008).

[60] P. Blöchl, Projector augmented-wave method, *Phys. Rev. B* **50**, 17953 (1994).

[61] G. Kresse and D. Joubert, From ultrasoft pseudopotentials to the projector augmented-wave method, *Phys. Rev. B* **59**, 1758 (1999).

[62] J. P. Perdew, K. Burke, and M. Ernzerhof, Generalized gradient approximation made simple, *Phys. Rev. Lett.* **77**, 3865 (1996).

[63] N. Marzari, A. A. Mostofi, J. R. Yates, I. Souza, and D. Vanderbilt, Maximally localized Wannier functions: Theory and applications, *Rev. Mod. Phys*. **84,** 1419 (2012).

[64] A. A. Mostofi, J. R. Yates, G. Pizzi, Y. S. Lee, I. Souza, D. Vanderbilt, N. Marzari, An updated version of wannier90: A tool for obtaining maximally-localised Wannier functions**,** *Comput. Phys. Commun*. **185**, 2309 (2014)

[65] Q. S. Wu, S. N. Zhang, H. -F. Song, M. Troyer, and A. A. Soluyanov, WannierTools: An open-source software package for novel topological materials, *Comput. Phys. Commun*.**224**, 405 (2018).

[66] J. R. Yates, X. Wang, D. Vanderbilt, and I. Souza, Spectral and Fermi surface properties from Wannier interpolation, *Phys. Rev. B* **75**, 195121 (2007).

[67] J. R. Yates, X. Wang, D. Vanderbilt, and I. Souza, Spectral and Fermi surface properties from Wannier interpolation, *Phys. Rev. B* **75**, 195121 (2007).

[68] T. Williams, C. Kelley, H. B. Broker, J. Campbell, R. Cunningham, D. Denholm, E. Elber, R. Fearick, C. Grammes, and L. Hart, Gnuplot 4.5: An interactive plotting program, 2011. *URL http://www. gnuplot. Info* (2017).

[69] M. A. Caprio, LevelScheme: A level scheme drawing and scientific figure preparation system for Mathematica. *Comp. Phys. Commun.* **171**, 107 (2005).